\PassOptionsToPackage{dvipsnames}{xcolor}

\documentclass[sigconf,10pt,screen]{acmart}
\AtBeginDocument{%
  \providecommand\BibTeX{{%
    \normalfont B\kern-0.5em{\scshape i\kern-0.25em b}\kern-0.8em\TeX}}}

\setcopyright{none}

\acmConference[]{}{}{}
\acmYear{}
\acmISBN{} 
\acmDOI{} 

\usepackage{url}
\usepackage{listings}
\usepackage{multirow}

\newcommand{\REM}[1]{}

\usepackage{fancyvrb}
\usepackage{float}
\usepackage{hyperref}
\usepackage{booktabs}
\usepackage{multirow}
\newsavebox{\codebox}
\usepackage[caption=false]{subfig}
\usepackage{backnaur}
\usepackage{xcolor}
\usepackage{pgfplots} 
\usetikzlibrary{patterns}

\usepackage{comment}

\begin{document}

\title[Parallel Computing for Medical Photoacoustic Image Reconstruction]{\huge GPU-Based Parallel Computing Methods for Medical Photoacoustic Image Reconstruction}

\author{Xinyao Yi}
\email{xyi2@uncc.edu}
\affiliation{%
  \institution{University of North Carolina at Charlotte}
  \city{Charlotte}
  \state{North Carolina}
  \country{USA}
  \postcode{28223}
}

\author{Yuxin Qiao}
\email{yq83@nau.edu}
\affiliation{%
  \institution{Northern Arizona University}
  \city{Flagstaff}
  \state{Arizona}
  \country{USA}
  \postcode{86011}
}

\renewcommand{\shortauthors}{Anjia Wang, et al.}

\begin{abstract}
Recent years have witnessed a rapid advancement in GPU technology, establishing it as a formidable high-performance parallel computing technology with superior floating-point computational capabilities compared to traditional CPUs. This paper explores the application of this technology in the field of photoacoustic imaging, an emerging non-destructive testing technique in biomedical engineering characterized by its high contrast, resolution, and penetration depth. We conduct a data parallelism analysis targeting the computationally intensive image reconstruction segment of photoacoustic imaging. By parallelizing the serial code for iterative reconstruction and optimizing memory access, we achieve significant improvements in processing speed. Our experiments compare the imaging speeds of vascular images reconstructed using CPUs and GPUs, with the results visualized using Matlab. The findings demonstrate that, while maintaining data accuracy, GPU parallel computing methods can markedly accelerate photoacoustic image reconstruction. This acceleration has the potential to facilitate the broader adoption of photoacoustic imaging in applications such as hemodynamic monitoring, clinical disease diagnosis, and drug development.
\end{abstract}


%


\maketitle

\section{Introduction}
\label{sec:intro}
As science and technology advance, we encounter increasingly complex graphics computing problems and significant data machine losses~\cite{chen2014data}. Concurrently, the computational demands in scientific fields have escalated, calling for enhanced high-performance computing solutions~\cite{cano2018survey, eklund2013medical}. Graphics Processing Units (GPUs), designed specifically for image computations, comprise numerous computing units, offering a rapid parallel computing framework for large-scale data processing with superior floating-point capabilities.

In recent years, GPU parallel computing technology has seen rapid development and widespread application across various domains, primarily due to its cost-effectiveness and efficiency in handling large-scale computations~\cite{yoon2020parallel}. One such application is in the field of photoacoustic imaging, an emerging non-destructive testing technique in biomedical engineering. Photoacoustic imaging is distinguished by its high contrast, resolution, and penetration depth, making it a promising medical imaging modality with significant potential for early diagnosis of cancer and cardiovascular and cerebrovascular diseases~\cite{yang2021photoacoustic}.

However, the vast data volume and complexity of photoacoustic image processing pose challenges that exceed the capabilities of traditional CPUs~\cite{lutzweiler2013optoacoustic}, thereby impeding the progress of medical photoacoustic imaging. Addressing this issue requires efficient medical image processing and analysis techniques to accurately and swiftly analyze massive image datasets. The rapid evolution of GPU parallel computing offers a promising solution to this challenge by leveraging the immense computational power of GPUs for handling the characteristics of massive data, complex algorithms, and high parallelism inherent in medical images.

This paper delves into GPU parallel computing methods, focusing on the parallel acceleration of iterative reconstruction algorithms during the image reconstruction process. By significantly enhancing the reconstruction speed, this study lays the groundwork for the future realization of real-time iterative reconstruction of photoacoustic images, thereby advancing the field of medical imaging.

\section{Parallel Computing on GPU}
\label{sec:motivation}
\subsection{Parallel Computing on GPU}

\subsubsection{Emergence and Development of GPU Technology}

The Graphics Processing Unit (GPU) stands as a pivotal element of a graphics card, predominantly dictating its performance tier and capabilities. In 1999, NVIDIA Corporation released the GeForce 256 graphics processor, which established the conceptual foundation of the modern GPU. This pivotal innovation ignited a demand for increasingly sophisticated applications, propelling a sustained and dynamic expansion of the industry. Through iterative advancements, the programmability of GPUs has been progressively augmented, leading to their utilization across a diversifying array of computational tasks.

In the GPU's nascent phase before 1999, programmability was notably absent, confining their role to hardware-accelerated image rendering~\cite{peddie2023gpu}. The advent of the second-generation GPU, spanning from 1999 to 2002, witnessed a decentralization of functionalities from the CPU, thereby broadening the application spectrum of GPUs. The year 2001 marked a significant milestone with NVIDIA’s GeForce3 and ATI’s Radeon 8500, which redefined graphics pipelines as stream processors, endowing them with vertex-level programmability and, albeit to a lesser extent, pixel-level programmability~\cite{mcclanahan2010history}.

The post-2002 era heralded the third-generation GPUs characterized by a substantial reduction in the complexity associated with general-purpose computational programming. A paradigm shift occurred in 2007 with NVIDIA’s introduction of the CUDA (Compute Unified Device Architecture) programming environment. This development metamorphosed the GPU into a versatile, programmable, high-performance parallel computing platform, inaugurating a new era dedicated to general-purpose computational applications.

\subsubsection{Computational Advantages of GPUs}

\begin{figure}[!htb]
\centering
\includegraphics[width=0.9\linewidth]{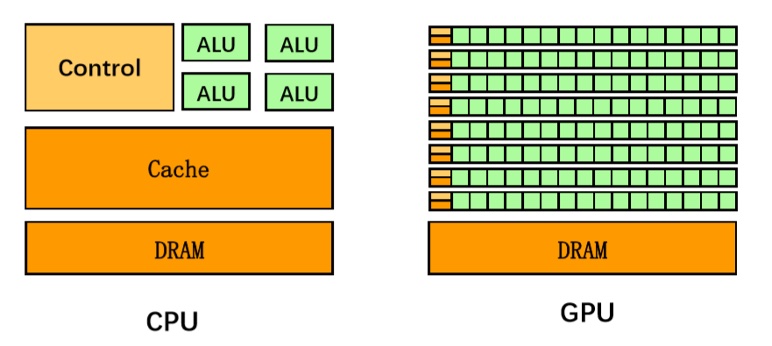}
\caption{CPU and GPU Structures}
\label{fig:gpu}
\end{figure}

The computational superiority of GPUs over CPUs stems from the inherent differences in their design philosophies and operational objectives. Figure~\ref{fig:gpu} illustrates the architectures of a CPU and a GPU. In a CPU, a significant portion of the chip (approximately 70\%) is dedicated to building multi-level caches and control units related to instruction-level parallelism, including out-of-order execution, branch prediction, and other logic-related components. Conversely, the arithmetic units occupy a relatively smaller fraction of the chip. On the other hand, a GPU is essentially a massive array of computing units, with a greater emphasis on arithmetic units, including integer and floating-point multiply-add units, and specialized computing units.

Moreover, unlike CPUs, GPUs possess higher-bandwidth memory, commonly referred to as video memory (VRAM), which enables them to perform exceptionally well in applications requiring high throughput. Consequently, while CPUs excel at handling instructions with complex logical interdependencies, GPUs have a distinct advantage in processing data that can be executed in parallel without dependencies.

\subsubsection{GPU+CPU Heterogeneous Parallel Computing Model}

The heterogeneous computing model, which synergistically combines GPUs with CPUs, capitalizes on their complementary processing strengths. Initially designed for high-throughput graphics rendering, GPUs possess inherent parallel processing capabilities that have significantly evolved beyond their original scope. The convergence of enhanced programmability and computational prowess has extended GPU applications to a broad spectrum of computational tasks.

In this integrated model, the serial and complex logical processing tasks are adeptly handled by the CPU, leveraging its optimized cores for sequential execution. Conversely, the GPU is employed for its extensive array of compute units, adept at parallel processing large-scale data. Such a division of labor enables each processor to function in its area of competence, thereby maximizing overall system efficiency. This collaborative approach can yield computational performance gains by orders of magnitude over CPU-only systems.

In 2007, NVIDIA Corporation developed the Compute Unified Device Architecture (CUDA), a paradigm-shifting framework that catalyzed the widespread adoption and growth of GPU parallel computing~\cite{demir2010compute}. As an extension of the C language, CUDA enables programming of GPU code using standard C, rendering the code compatible with both CPUs and GPUs. This breakthrough eliminated the need for developers to master complex shading languages or graphical primitives for GPU programming. Consequently, a model for heterogeneous hybrid parallel computing combining GPU and CPU was proposed.

The combination of CPU and GPU harnesses the strengths of both processors: the CPU, with its cores optimized for sequential processing, handles the serial components of a program, while the GPU, comprising thousands of compute units, manages the parallel tasks. This collaborative computational model significantly enhances the capabilities of both CPUs and GPUs, achieving efficiency gains that can surpass traditional CPU-only computations by several orders of magnitude.

\begin{figure}[h]
\centering
\includegraphics[width=\linewidth]{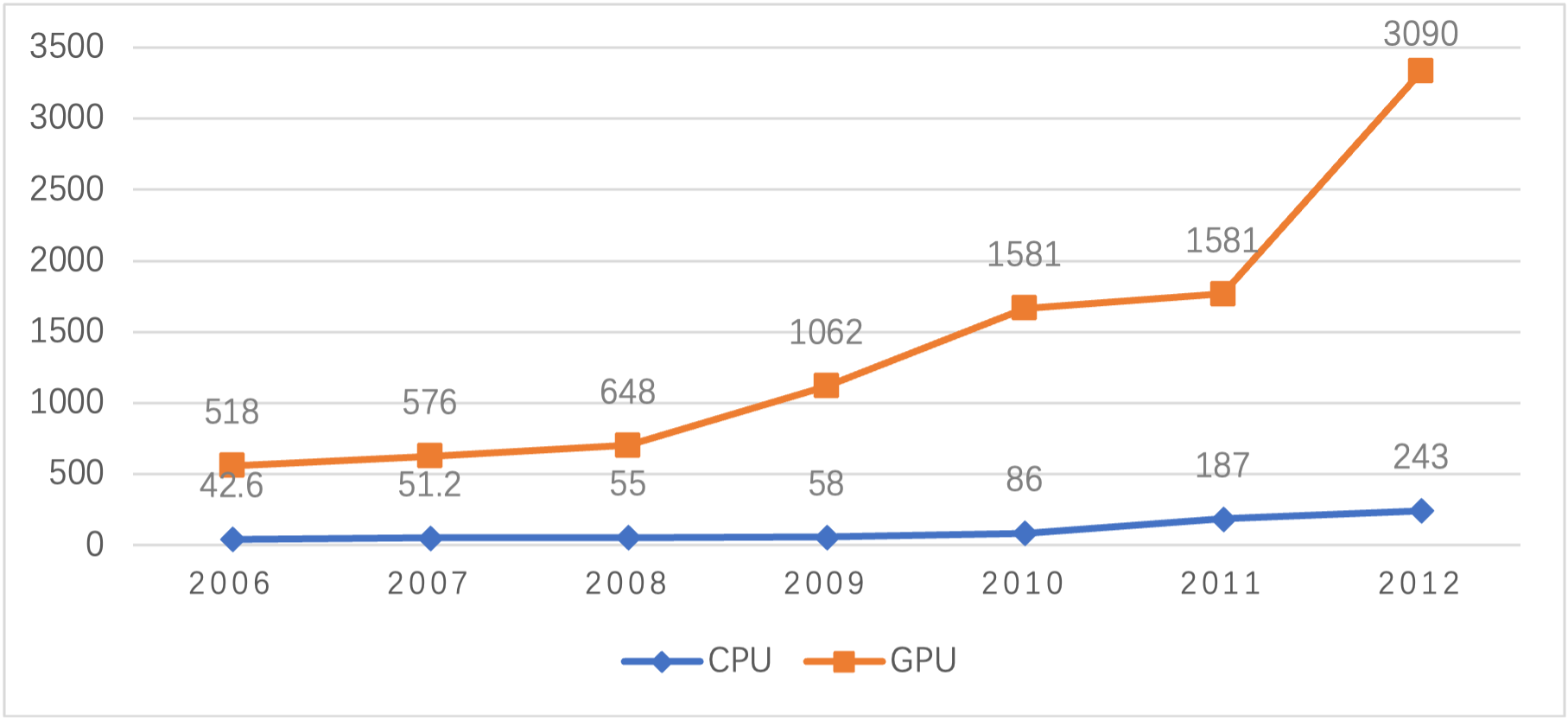}
\caption{Peak performance of CPU and GPU (measured in billions of floating-point operations per second, or gigaflops)}
\label{fig:gpu2}
\end{figure}

As depicted in Figure \ref{fig:gpu2}, the gap in computational capabilities between CPUs and GPUs has widened with technological advancements. By 2012, the speed difference had already reached approximately 12-fold. This disparity underscores the rapid pace at which GPU technology has evolved, significantly outpacing the improvements in CPU performance. The increasing parallel processing power of GPUs makes them an ideal choice for tasks requiring high computational throughput, further solidifying their advantage in various applications, including medical photoacoustic imaging.

\subsection{CUDA}
Since its introduction, CUDA (Compute Unified Device Architecture) has experienced rapid development, becoming a cornerstone in leveraging the power of GPUs. CUDA allows developers to efficiently utilize GPUs using mainstream programming languages such as C, C++, and Fortran, without the need to learn new languages. This accessibility has significantly lowered the barrier to entry for GPU programming, enabling a wider range of researchers and developers to harness the parallel computing capabilities of GPUs for various applications, including medical image processing and scientific computations. CUDA's comprehensive ecosystem, which includes libraries, development tools, and support forums, further facilitates the development and optimization of GPU-accelerated applications.

\subsubsection{Programming Model}

Within the CUDA architecture, the CPU is designated as the host and the GPU as the device, with both working in conjunction to perform computations. CUDA code is bifurcated into two segments: standard C code executed on the CPU and parallel code, known as a kernel, executed on the GPU and compiled using NVIDIA's nvcc compiler.

\begin{figure}[h]
\centering
\includegraphics[width=\linewidth]{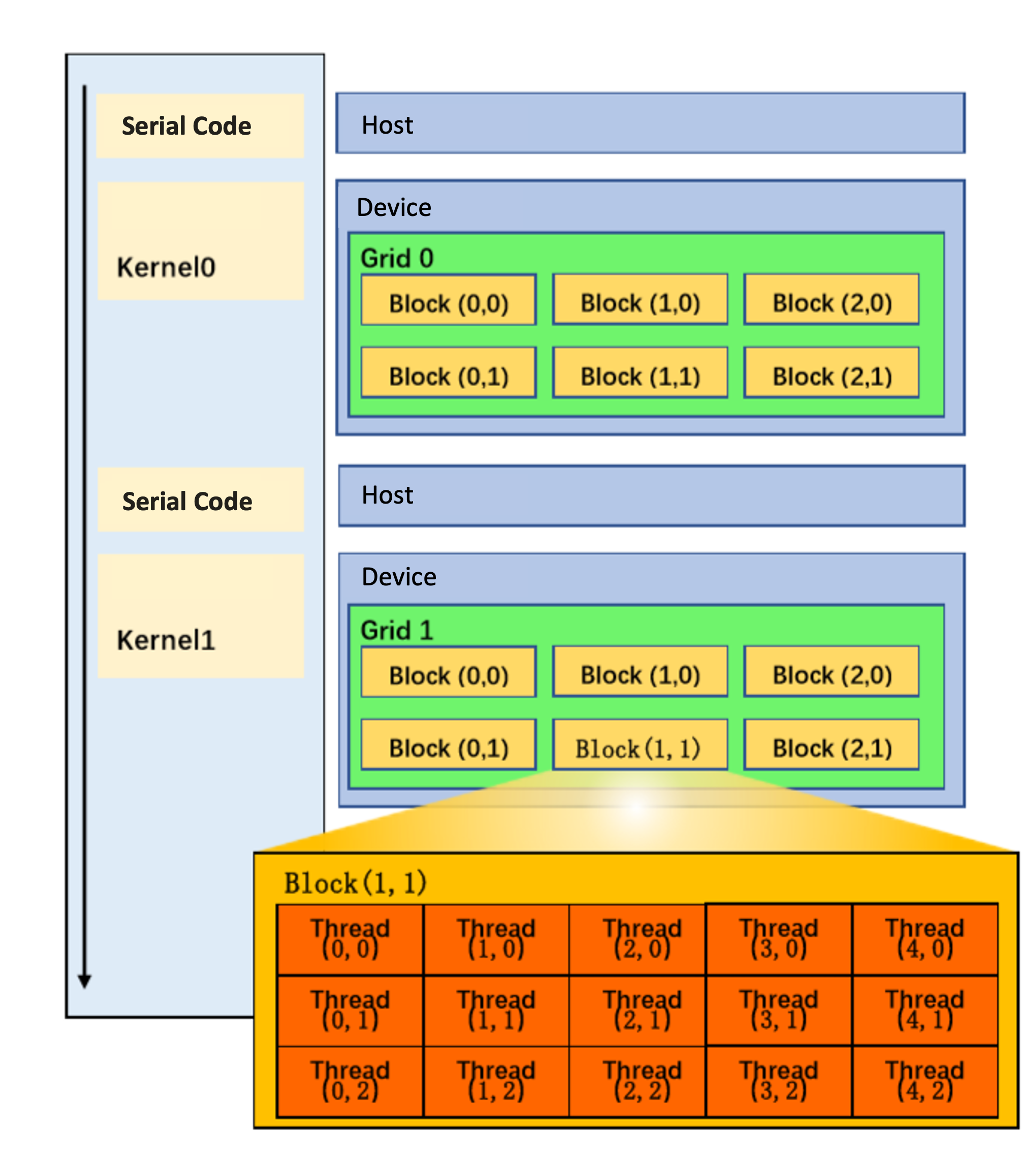}
\caption{CUDA Programming Model}
\label{fig:Programming}
\end{figure}

Figure \ref{fig:Programming} illustrates the operational principle of the GPU under the CUDA framework. On the GPU, invoking a kernel function initiates a grid, with each kernel corresponding to a grid. Each grid is composed of one or more thread blocks (blocks), executed on a Streaming Multiprocessor (SM); within each thread block, there are multiple threads (threads) that run on the SM's Compute Units (SPs).

Every thread block is assigned a two-dimensional identifier via CUDA's specific keywords blockIdx.x and blockIdx.y. All thread blocks are organized in the same manner and contain an equal number of threads. There are two levels of parallelism during execution: blocks within the grid operate in parallel, as do threads within each block. While thread blocks cannot communicate with one another, threads within a single block can communicate and synchronize via shared memory (shared memory).

\subsubsection{Memory Model}
In CUDA, distinct memory spaces exist for the host and the device. Consequently, when executing a kernel on the device, programmers must transfer data from host memory to the allocated device memory space. Subsequently, upon completion of kernel execution on the device, results must be transferred back to the host, and the device's memory space must be released.

\begin{figure}[h]
\centering
\includegraphics[width=\linewidth]{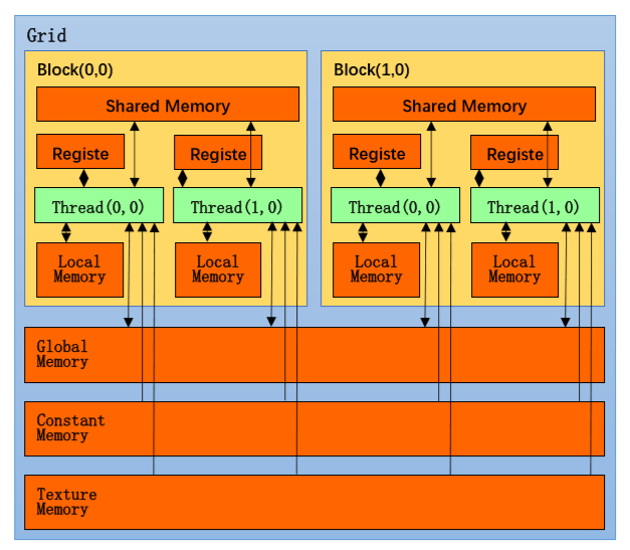}
\caption{GPU Hydrogenous Memory Architecture}
\label{fig:memory}
\end{figure}

Figure \ref{fig:memory} illustrates the memory model of a GPU, encompassing various memory types. Each thread possesses private registers and local memory. Shared memory is present within each thread block, facilitating communication among threads within the same block. Global memory resides off-chip on the GPU board and is accessible for reading and writing by all threads in a grid. Constant memory and texture memory are restricted to reading operations by threads. Register access incurs nearly zero latency, and accessing shared memory without bank conflicts has latency comparable to register access. In contrast, global memory in the GPU experiences higher latency, approximately 400-800 clock cycles. Consequently, when designing a program, it is advisable to utilize shared memory for data storage to reduce data retrieval time and enhance overall program execution speed. Notably, texture memory is well-suited for image processing operations due to its ability to accelerate access to large amounts of data with random and non-aligned access patterns.

Memory access latency constitutes a significant limiting factor for GPU processing speed~\cite{yi2021cudamicrobench}. Effective memory planning is crucial when designing a program to achieve maximum access bandwidth. Various methods, such as switching between different blocks on a Streaming Multiprocessor (SM), storing multiple block contexts in an SM, and scheduling another block for execution on the core when one block is suspended due to latency, can be employed to hide memory access latency~\cite{yi2023neorodinia}.

\subsubsection{Integration of CUDA with MATLAB for Enhanced Parallel Computing}

MATLAB demonstrates excellent performance in matrix operations, plotting functions and data, and interfacing with other programming languages. It is primarily used in fields such as engineering computation, control design, signal processing, and communications. With the increasing scale of data computation, many laboratories and research institutions are turning to parallel computing to improve computational speed. MATLAB provides a Parallel Computing Toolbox that can leverage GPUs to enhance code execution speed. However, the Parallel Computing Toolbox has various constraints and limitations, preventing GPUs from fully leveraging their high-performance advantages and resources from being utilized to their full potential. Therefore, combining CUDA with MATLAB programming provides an excellent approach to problem-solving. Assignments, conditional structures, and small-scale matrix operations can be handled by MATLAB, while CUDA can be used for large-scale, time-consuming matrix operations. During this process, MATLAB's Parallel Computing Toolbox can also assist in performing some small-scale matrix operations, resulting in the best acceleration effect for the program.

\begin{figure}[h]
\centering
\includegraphics[width=\linewidth]{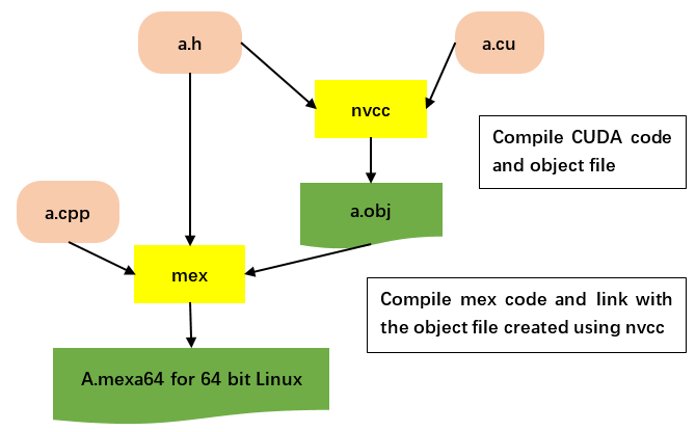}
\caption{Compilation process of kernel function with MATLAB}
\label{fig:matlab}
\end{figure}

In MATLAB, the CUDA compilation process is illustrated in Figure \ref{fig:matlab}. Firstly, three files need to be compiled in the CUDA model, namely a.h, a.cu, and a.cpp files (where "a" represents the function to be compiled). Using the CUDA built-in compiler nvcc, the a.h and a.cu files are compiled into an a.obj file. After compiling the a.obj file, the mex compiler in MATLAB is used to compile the a.cpp and a.obj files into a callable function A. Once the compilation is complete, function A can be directly called in MATLAB.

\section{Photoacoustic Imaging}
\label{sec:upir-parallelism}
\subsection{Introduction to Medical Photoacoustic Imaging}

The domain of biomedical imaging has experienced remarkable advancement since the late 19th century. During this period, various imaging modalities emerged, yet many were either exorbitantly costly or posed significant risks to human health~\cite{fujimoto2000optical}. Over the last two decades, the advent of photoacoustic imaging has addressed these challenges. This modality synergizes the high contrast of optical imaging with the deep penetration capabilities of ultrasound imaging, offering a cost-effective, high-quality, and non-invasive imaging solution that has received widespread recognition~\cite{walker2016species}.

The foundational concept of photoacoustic imaging traces back to the discovery of the photoacoustic effect by Alexander Graham Bell in 1880~\cite{anderson1981optics}. The subsequent progress in laser technology during the 1960s propelled the application of photoacoustic techniques in industrial and scientific domains. By the late 20th century, this technology was harnessed for biomedical imaging, culminating in the generation of the first photoacoustic medical images. Since then, the evolution of photoacoustic imaging technology has been swift. In comparison to conventional ultrasound imaging, it offers superior optical contrast and provides critical physiological parameters such as hemoglobin oxygen saturation. Moreover, it achieves enhanced spatial resolution for deep tissue imaging relative to pure optical imaging modalities. Despite its relatively recent inception, photoacoustic imaging has been extensively applied in various preclinical imaging scenarios, attracting significant attention~\cite{ku2005imaging,zhang2006functional}. This technique typically utilizes unfocused ultrasound transducers with mechanical scanning or ultrasound transducer arrays for image data acquisition~\cite{wang2003noninvasive,li2010real}, followed by the application of reconstruction algorithms to retrieve the optical properties of the imaged object.

\begin{figure}[h]
\centering
\includegraphics[width=0.5\textwidth]{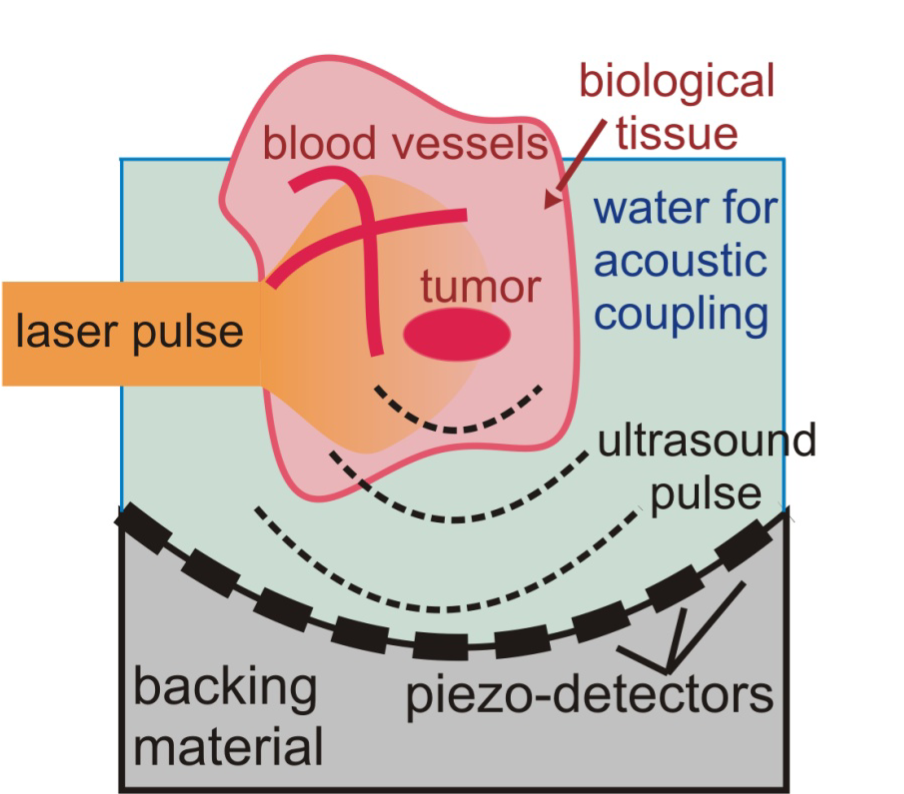}
\caption{Principle diagram of the photoacoustic effect.}
\label{fig:principle}
\end{figure}

The core principle of biomedical photoacoustic imaging is illustrated in Figure \ref{fig:principle}. A wide-beam, short-pulse laser irradiates biological tissue, leading to the absorption of light energy and subsequent thermal expansion, which generates pressure waves and produces the photoacoustic signal. Ultrasound detectors placed on the surface of the tissue capture these emitted ultrasound waves. By analyzing the detected photoacoustic signals, the distribution of light energy absorption within the tissue is reconstructed to create an image. The differential light energy absorption properties of various biological tissues enable photoacoustic imaging to produce high-contrast structural images. Additionally, when integrated with photoacoustic spectroscopy, this technique facilitates the quantitative assessment of changes in tissue components and offers a comprehensive depiction of minute tissue lesions and crucial physiological parameters, thereby enhancing functional imaging. These unique advantages render photoacoustic imaging particularly valuable for the early diagnosis of conditions such as breast cancer and prostate cancer.

\subsection{Advancements and Applications of Medical Photoacoustic Imaging}
Medical photoacoustic imaging has become an indispensable part of modern medicine, with numerous scholars at home and abroad conducting in-depth research on this technology~\cite{wang2003noninvasive,li2010real}. In March 2014, the research group led by Liang Song at the Institute of Biomedical and Health Engineering, Shenzhen Institutes of Advanced Technology, Chinese Academy of Sciences, developed an intravascular optical-resolution photoacoustic tomography (OR-PAT) system with an imaging resolution of up to 19.6 micrometers. This technology can provide information reflecting important physiological functions such as tissue composition and inflammation activity, playing a significant role in guiding the interventional diagnosis and treatment of cardiovascular diseases. 
\begin{figure}[h]
\centering
\includegraphics[width=0.5\textwidth]{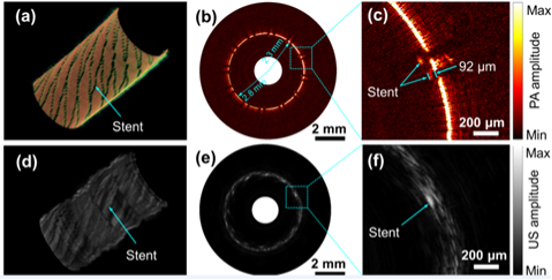}
\caption{Evaluation of stent apposition using high-resolution intravascular photoacoustic microscopy (a–c) and intravascular ultrasound imaging (d–f).}
\label{fig:related1}
\end{figure}

In March 2015, Dr. Lihong Wang and his research team at Washington University in St. Louis developed a fast-functional photoacoustic microscopy (PAM) technique that can detect blood flow, blood oxygen, oxygen metabolism, and other functions in the living mouse brain.

\begin{figure}[h]
\centering
\includegraphics[width=0.5\textwidth]{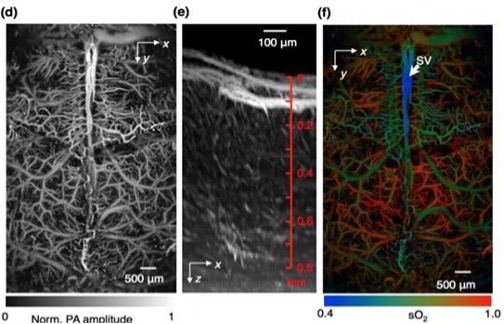}
\caption{Images of mouse brain captured by fast-functional photoacoustic microscopy technology: (d) is the vascular system within the entire skull projected onto the two-dimensional x-y plane; (e) is an enhanced imaging image of a typical brain vascular system projected onto the two-dimensional x-z plane; (f) is an image of mouse brain hemoglobin oxygen saturation captured by photoacoustic microscopy, taken using a new method based on single wavelength and pulse width with two laser beams.}
\label{fig:related2}
\end{figure}

\subsection{Principle of Photoacoustic Computed Tomography (PACT)}

Photoacoustic computed tomography (PACT) allows the acquisition of photoacoustic images. According to the theory of photoacoustic generation \cite{ref17}, the acoustic pressure $p(\mathbf{r},t)$ at position $\mathbf{r}$ and time $t$ satisfies the wave equation:
\begin{equation}
    \nabla^2 p(\mathbf{r},t) - \frac{1}{c^2} \frac{\partial^2 p(\mathbf{r},t)}{\partial t^2} = \frac{\partial^2 p_0(\mathbf{r})}{\partial t^2}, \quad
\end{equation}
where $p_0(\mathbf{r})$ is the initial photoacoustic pressure induced by the laser or electromagnetic pulse, and $c$ is the speed of light. By solving the wave equation, the forward model using prediction can be written as:
\begin{equation}
    p(\mathbf{r},t) = \text{FWD}[p_0(\mathbf{r})], \quad
\end{equation}
where $\text{FWD}$ represents the forward operator, and $p_0(\mathbf{r})$ is the initial pressure distribution.

The time-dependent wave equation can be transformed into the time-frequency domain by taking the Fourier transform of $p(\mathbf{r},t)$ to obtain $P(\mathbf{r},\omega)$, where $\omega$ is the angular frequency:
\begin{equation}
    \nabla^2 P(\mathbf{r},\omega) + \frac{\omega^2}{c^2} P(\mathbf{r},\omega) = -\omega^2 P_0(\mathbf{r}), \quad
\end{equation}
where $P_0(\mathbf{r})$ is the Fourier transform of the initial pressure $p_0(\mathbf{r})$, and $\omega$ is equal to the angular frequency corresponding to the signal's time frequency. Based on equations (2) and (3), the time-frequency domain forward model can be expressed as:
\begin{equation}
    P(\mathbf{r},\omega) = \text{FWD}_\omega[P_0(\mathbf{r})], \quad
\end{equation}
where $\text{FWD}_\omega$ represents the forward operator in the frequency domain.

\subsection{The Principle of Iterative Reconstruction for Photoacoustic Imaging}

Photoacoustic image reconstruction is a core part of photoacoustic imaging technology. The quality of the reconstruction method directly affects the quality of the photoacoustic image and the speed of image acquisition. The most common reconstruction method currently is the Back-Projection (BP) algorithm, which has high resolution and relatively fast imaging speed. However, it has certain limitations in practical applications. When the projection data is unevenly distributed, sparse, or when there is significant noise, the BP method's reconstruction effect is poor, leading to significant artifacts in the results. Thus, Iterative Reconstruction (IR) methods have emerged. Iterative reconstruction algorithms are suitable for reconstructing high-quality images from fewer projection data. However, algebraic iterative reconstruction algorithms are slow because they require multiple iterations to converge and require large data storage, making their lengthy processing time the biggest obstacle to their application. Therefore, improving the speed of iterative reconstruction has important practical value.

\subsubsection{Compressed Sensing}
Compressed Sensing (CS) is a method for recovering sparse signals from undersampled measurements, widely used in CT and Diffuse Optical Tomography. Numerous studies have certified that reconstruction techniques based on CS can recover photoacoustic signals at sampling rates lower than what is demanded by the Nyquist sampling theorem.

The prerequisite is that the signal or its transform in a certain domain is sparse or compressible, and by finding an appropriate sparse transform \( x = \psi\theta \), most medical images are sparse in some domain.  \( \theta \) represents the original image, and \( x \) represents the transformed image. It has been proven that photoacoustic images can be transformed into a sparse domain through various transformations, such as numerical derivatives and wavelet transforms.

In PACT, the objective of reconstruction is to recover the photoacoustic signal \( \theta \) from the measurements \( y \) obtained by the probe, and if the measurement matrix \( K \) obtains \( y \), then \( y = K\theta \). Photoacoustic images with CS can typically be reconstructed by solving the constrained optimization problem.
\begin{equation}
\min \|x\|_1 \quad \text{s.t.} \quad y = K\psi^{-1}x.
\end{equation}

To implement the CS reconstruction of PACT, the discrete expressions of the PACT system in the time and frequency domains for \( K \) will be given out below.
Time domain measurement matrix:
\begin{equation}
\begin{aligned}
K(m,t)_{(i,j)} = \frac{1}{2\pi c} \delta\left(t- \frac{|r_{i,j} - r_m|}{c} \right), \\
\quad m=1,2,\dots,p; t=s\Delta t, s=1,2,\dots,q_s.
\end{aligned}
\end{equation}

Frequency-domain reconstruction:

\begin{equation}
\begin{aligned}
K(m,n)_{(i,j)} = ick_n \frac{\exp(-ik_n |r_{i,j} - r_m|)}{|r_{i,j} - r_m|}, \\
\quad m=1,2,\dots,p; n=1,2,\dots,q_n.
\end{aligned}
\end{equation}

Wherein, \( r_{i,j} \) represents the Cartesian coordinates of the image pixels, \( r_m \) represents the position of the probe, \( p \) represents the number of probes, \( q_s \) represents the number of sampling points in the time domain, \( q_n \) represents the number of sampling points in the frequency domain.

\subsubsection{Reconstruction Method}
Based on the above CS theory, the CS-based reconstruction model for PACT is represented by the following formula:
\begin{equation}
\arg\min_x F = \| \Phi x - y \|_2^2 + \alpha \| x \|_1 + \beta \mathrm{TV}(\Psi^\prime x)
\end{equation}

\( F \) is an objective function composed of three parts. The first part represents the squared error between the estimated measurement values of the reconstructed signal and the experimental measurement values. The second part represents the \( \ell_1 \) norm of the signal in the sparse domain. The third part represents the total variation (TV) penalty of the signal. \( \alpha \) and \( \beta \) are regularization parameters that determine the trade-off between data consistency and sparsity.

\section{Implementation}
\label{sec:upir-data}
The experimental evaluation contains two distinct segments. The initial segment encompasses a comprehensive assessment of image reconstruction algorithms, specifically juxtaposing back-projection and iterative methodologies. This evaluation rigorously examines the comparative imaging quality and temporal efficiency inherent to each algorithm, thereby elucidating the respective merits and limitations of the time-reversal reconstruction approach via a meticulous comparative analysis.

The subsequent segment delves into the application of GPU parallel computing within the iterative reconstruction framework, as applied to angiographic imaging of murine vasculature. This involves a strategic enhancement of the conventional serial reconstruction paradigm, wherein parallel optimization techniques are employed to significantly expedite the computational process.

The computational hardware employed in this study consists of a CPU model Intel(R) Xeon(R) CPU E5-2620 v2 @2.10GHz, featuring a dual physical GPU setup, with a memory capacity of 64GB, and a substantial storage space of 6TB.

The graphics processing unit (GPU) utilized for the research is the NVIDIA-produced Tesla K40c, built on the NVIDIA Kepler™ architecture, and recognized as a formidable contender in high-performance computing cards. The attributes of the GPU, as empirically determined, are delineated in Table \ref{tab:table1}. The Tesla K40c GPU boasts a computational capability of 3.5, with a global memory of 12GBytes, a constant memory provision of 65536 bytes, and an allowance of a maximum shared memory of 49152 bytes within each thread block. The GTX960 series permits the configuration of the grid to the maximal dimensions of (2147483647, 65535, 65535), where the maximum number of threads allowed per thread block is set at 1024.

\begin{table}[]
\centering
\begin{tabular}{|l|l|}
\hline
GPU Model & Tesla K40c \\
\hline
CUDA Version & 8.0 \\
\hline
Compute Capability & 3.5 \\
\hline
Memory & 12 GBytes \\
\hline
Number of CUDA Cores & 2880 \\
\hline
Constant Memory & 65536 bytes \\
\hline
Shared Memory per Block & 49152 bytes \\
\hline
Texture Memory & 512 bytes \\
\hline
\end{tabular}
\caption{Tesla K40c GPU Configuration}
\label{tab:table1}
\end{table}

\subsection{Experiment 1}
In this experiment, photoacoustic imaging of murine vascular structures was reconstructed and analyzed. Comparative evaluations were conducted between back-projection algorithms and iterative reconstruction algorithms to ascertain their imaging efficacy and temporal efficiency. This analysis enabled the identification of the strengths and limitations inherent in the iterative approach.

\begin{figure}[h]
\centering
\includegraphics[width=0.45\textwidth]{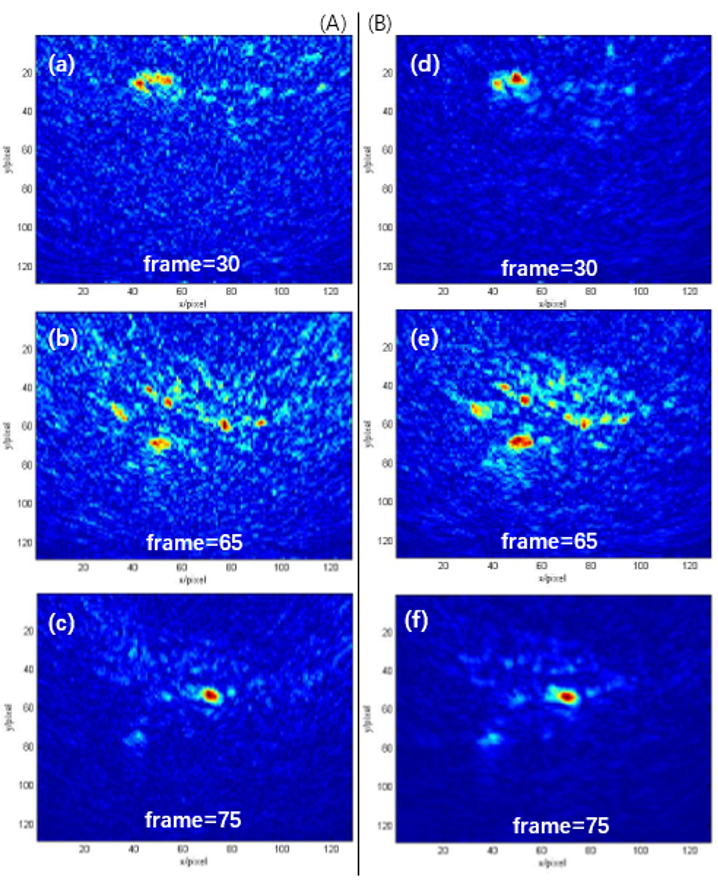}
\caption{Comparison of filtered back-projection reconstruction and time-reversal reconstruction of mouse vascular photoacoustic images. (A) Mouse vascular photoacoustic image reconstructed with the back-projection algorithm; (B) Mouse vascular photoacoustic image reconstructed with the time-reversal reconstruction algorithm.}
\label{fig:exp1}
\end{figure}

The photoacoustic imaging system captured 166 frames of murine vascular imaging. Due to the large size of the images and consequent slow processing speeds, three frames were selected for the experiment. For each frame, a segment of 127 pixels by 127 pixels was extracted for the reconstruction trials. As illustrated in Figure \ref{fig:exp1}, Section (A) displays the reconstructed images of the 30th, 65th, and 75th frames using the back-projection algorithm. Corresponding images in Section (B) were reconstructed using the iterative algorithm. The reconstruction times are presented in Table \ref{tab:table2}.

\begin{table}[]
\centering
\begin{tabular}{|l|c|c|c|}
\hline
Method & \multicolumn{3}{c|}{Reconstruction Time (s)} \\
\hline
 & frame=30 & frame=65 & frame=75 \\
\hline
BP & 2.277 & 2.712 & 2.666 \\
\hline
IR & 118.470 & 118.814 & 117.082 \\
\hline
\end{tabular}
\caption{Reconstruction times using Back-Projection (BP) and Iterative Reconstruction (IR) methods}
\label{tab:table2}
\end{table}

Analysis of Figure \ref{fig:exp1} reveals that, for the same frame, the iterative algorithm visually outperforms the back-projection algorithm. The latter's suboptimal reconstruction results are due to its non-exact solution-based approach, theoretically possessing inherent shortcomings: back-projection involves distributing weights along an arc, which may lead to non-zero grayscale values in the reconstructed image where the original image had zero values, producing artifacts that blur details and exacerbate with the complexity of the imaged object's structure. In contrast, the iterative method is less affected by such imaging artifacts and more accurately restores the original image.

However, as Table \ref{tab:table2} indicates, the iterative method has a significant drawback: the computational demand is enormous, and the reconstruction time for a single frame is excessively long. The data show that the time for iterative reconstruction is up to 50 times higher than that for back-projection, greatly limiting its application in photoacoustic image reconstruction. To enhance the speed of iterative reconstruction in photoacoustic imaging and expand its practical application, it is crucial to apply GPU-based parallel acceleration to the most time-consuming computations.

\subsection{Experiment 2}

\subsubsection{Execution Time Analysis}

In iterative reconstruction, the execution of each step depends on the result of the previous step. This dependency prevents the steps from being processed in parallel. Therefore, serial code is time-tested to identify the most time-consuming execution statements for parallel optimization, thereby achieving an overall optimization effect.

For a frame of 128x128 photoacoustic images, each part of the code was time-tested in MATLAB. The time consumption of the most time-consuming code lines and their respective percentage are shown in Table \ref{tab:table3}.

\begin{table*}[ht]
\centering
\begin{tabular}{|l|c|c|c|}
\hline
Code & Calls & Total Time(s) & \%Time \\
\hline
t1=2*kn'*(kn*mk1r-y); & 90 & 135.485 & 96.7\% \\
\hline
t1=-2*kn'*y; & 1 & 2.060 & 1.5\% \\
\hline
t2=(tv'*inv(wmr)*tv*mk1+mk1*(tv*inv(wml)*tv')); & 90 & 0.433 & 0.3\% \\
\hline
All other lines & & 2.177 & 1.5\% \\
\hline
\end{tabular}
\caption{Execution Time Analysis}
\label{tab:table3}
\end{table*}

During the iterative reconstruction process, a few specific lines of code consume a disproportionately large amount of time, with one line accounting for as much as 96.7\% of the total time. The execution time of the program largely depends on the execution rate of this line of code. Therefore, the variables involved in this statement are outputted to clarify their computation process. \( kn \) is a double matrix of size 6144x16384, \( kn' \) is its transpose, \( mk1r \) is a double matrix of size 16384x1, and \( y \) is a double matrix of size 6144x1.

It is evident that in the iterative process, the multiplication of large matrices consumes a significant amount of time, hence the use of CUDA for parallel acceleration of the multiplication process.

\subsubsection{Parallel Optimization Strategies}

Parallel Optimization Strategies in Program Execution

The optimization of CUDA-based programs fundamentally entails the parallelization of the most computationally intensive tasks to reduce overall execution time. Within the scope of this study, it was observed that matrix multiplication constituted 96.7\% of the runtime. Traditional matrix multiplication algorithms implemented in a serial context require three nested loops—wherein the two outer loops calculate the products of corresponding elements across the nth rows and columns, and the innermost loop aggregates these products to establish the definitive value at each specified position within the resultant matrix. The inherent sequentiality of this method necessitates the completion of one computational step before the subsequent one begins, leading to potential task blocking and increased latency in execution.

Addressing these inefficiencies, GPU parallelization techniques have been applied to dissect the computational workload, allocating the calculation of individual elements of the resultant matrix to discrete cores. This concurrent processing paradigm significantly reduces the need for iterative looping and expedites computation, thereby optimizing efficiency. Further, this paper presents an advanced data accumulation strategy that transcends the traditional cyclic approach. A tree-based accumulation technique is proposed to elevate parallelism and diminish temporal overheads. Under this schema, for a given set of n data points requiring accumulation, the initial phase involves the division of the workload among n/2 CUDA threads, each handling the addition of two data elements in a parallel fashion, resulting in n/2 intermediate sums. Subsequent phases halve the thread count iteratively, each time performing parallel pairwise summations until a singular accumulated value is derived. This innovative algorithm not only reduces iterative loop counts but also mitigates thread underutilization, demonstrating a leap forward in computational efficiency for intensive parallel tasks.

\begin{figure}[h]
\begin{Verbatim}[fontsize=\scriptsize, frame=single, numbers=left,  xleftmargin=0.04\linewidth]
// Serial Pseudocode
// Input matrices A, B and output matrix C, 
// all of dimensions i x j
Input: A(i, j), B(i, j)
Output: C(i, j)

Main() {
    // Iterate over the rows of matrices A and C
    For i = 0 to n-1 {
        // Iterate over the columns of matrices B and C
        For j = 0 to n-1 {
            C(i, j) = 0;
            For k = 0 to n-1 {
                C(i, j) = C(i, j) + A(i, k) * B(k, j);
            }
        }
    }
}
\end{Verbatim}
\begin{Verbatim}[fontsize=\scriptsize, frame=single, numbers=left,  xleftmargin=0.04\linewidth]
// Parallel Pseudocode
// Input matrices A, B and output matrix C, 
// all of dimensions i x j

Kernel 1:
Input: A(i, j), B(i, j)
Output: M(i, j)

1. bx = blockIdx.x, by = blockIdx.y, 
   tx = threadIdx.x, ty = threadIdx.y;
2. Set Tiles in the shared memory of the block;
3. For i = 0 to n-1 {
    For j = 0 to n-1 {
        M(i, j) = A(i, tx) * B(tx, j);
    }
}

Kernel 2:
Input: M(i, j)
Output: C(i, j)

// Accumulate the products in M
For k = 0 to n-1 {
    C(i, j) = C(i, j) + M(i, j);
}
\end{Verbatim}
\caption{Serial and Parallel Pseudocode of the Matrix Multiplication}
\label{fig:serial_parallel}
\end{figure}

In the parallelized matrix multiplication algorithm, `Kernel 1` is responsible for conducting element-wise multiplications between matrices A and B, with the intermediate products being stored within matrix M. Subsequently, `Kernel 2` undertakes the task of aggregating these products, ultimately rendering the final elements of matrix C. The assignment of tasks to the GPU's processing cores is orchestrated via block and thread indices (bx, by, tx, ty), which is a convention rooted in CUDA programming. This methodology facilitates the simultaneous execution of operations, leveraging the GPU's architecture for enhanced computational throughput. It should be noted that the provided pseudocode abstracts from the complexities inherent in actual CUDA development, such as the meticulous synchronization of threads and the strategic allocation of shared memory resources. These aspects, while omitted for brevity in the pseudocode, are crucial for the effective implementation of parallel algorithms on GPU platforms.

\subsubsection{Optimizing Memory Utilization in Matrix Operations}

In the advancement of GPU computational capabilities, the efficiency of memory access emerges as a predominant bottleneck impacting program execution times. The latency inherent in accessing global memory far exceeds that of shared or texture memory access. Consequently, program optimization increasingly focuses on maximizing the use of shared and texture memory while curtailing dependence on global memory.

\begin{figure}[h]
\centering
\includegraphics[width=0.45\textwidth]{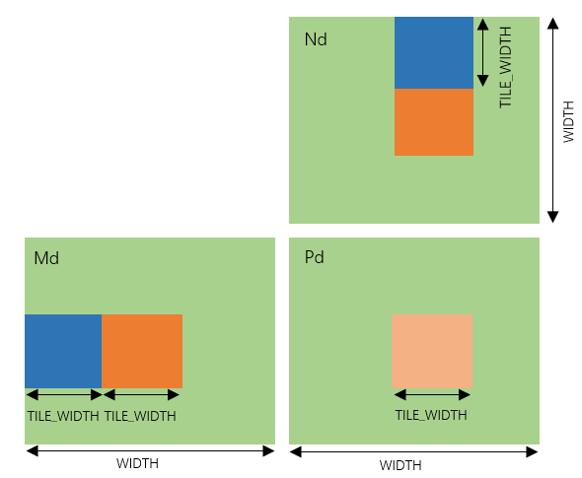}
\caption{Optimizing Memory Utilization in Matrix Operations}
\label{fig:storage}
\end{figure}

To accelerate computational throughput, a judicious strategy employed involves the segmentation of the dataset required for the calculation of specific portions of the resultant matrix, followed by the transference of this segmented data to shared memory. However, the finite capacity of shared memory dictates a need for meticulous data partitioning, which is exemplified in Figure \ref{fig:storage}. This data partitioning paradigm is instrumental in refining memory access patterns, which concomitantly enhances the efficiency of GPU-accelerated programs.

Within matrix operation contexts, where Nd and Md correspond to the input matrices and Pd to the output matrix, an adept strategy is to partition the extensive matrix into a series of tiles, effectively modularizing the program into several discrete computational stages. Each stage executes by performing accumulative operations on subsets of Md and Nd, thereby progressively assembling Pd. During these operations, data requisite for the computation of an individual tile is systematically migrated from global to shared memory. This stratagem optimizes data locality for each computational stage, significantly reducing the memory access latency and thereby bolstering the overall computational efficacy.

\subsubsection{Experimental Results}

To evaluate the computational advantages of GPUs relative to CPUs in photoacoustic image reconstruction involving matrix multiplication, this study initially performed a detailed analysis of the time performance associated with the matrix multiplication stages of the reconstruction process. The findings are systematically documented in Table \ref{tab:table4}.

\begin{table}[]
\begin{tabular}{|l|l|l|l|}
\hline
{\color[HTML]{0D0D0D} \textbf{Matrix A}} & {\color[HTML]{0D0D0D} \textbf{Matrix B}} & {\color[HTML]{0D0D0D} \textbf{CPU Time (s)}} & {\color[HTML]{0D0D0D} \textbf{GPU Time (s)}} \\ \hline
{\color[HTML]{0D0D0D} 6144x16384}                & {\color[HTML]{0D0D0D} 16384x1}                 & {\color[HTML]{0D0D0D} 42.3}                              & {\color[HTML]{0D0D0D} 7.8}                               \\ \hline
{\color[HTML]{0D0D0D} 16384x6144}                & {\color[HTML]{0D0D0D} 6144x1}                  & {\color[HTML]{0D0D0D} 38.4}                              & {\color[HTML]{0D0D0D} 6.4}                               \\ \hline
\end{tabular}
\caption{Time Consumption Comparison for Matrix Multiplication between GPU and CPU}
\label{tab:table4}
\end{table}

The comparison confirms that the result matrices generated by both the GPU and CPU are identical; however, the GPU's computational speed is approximately six times faster than that of the CPU, indicating a substantial enhancement in performance efficiency.

\begin{figure}[h]
\centering
\includegraphics[width=0.45\textwidth]{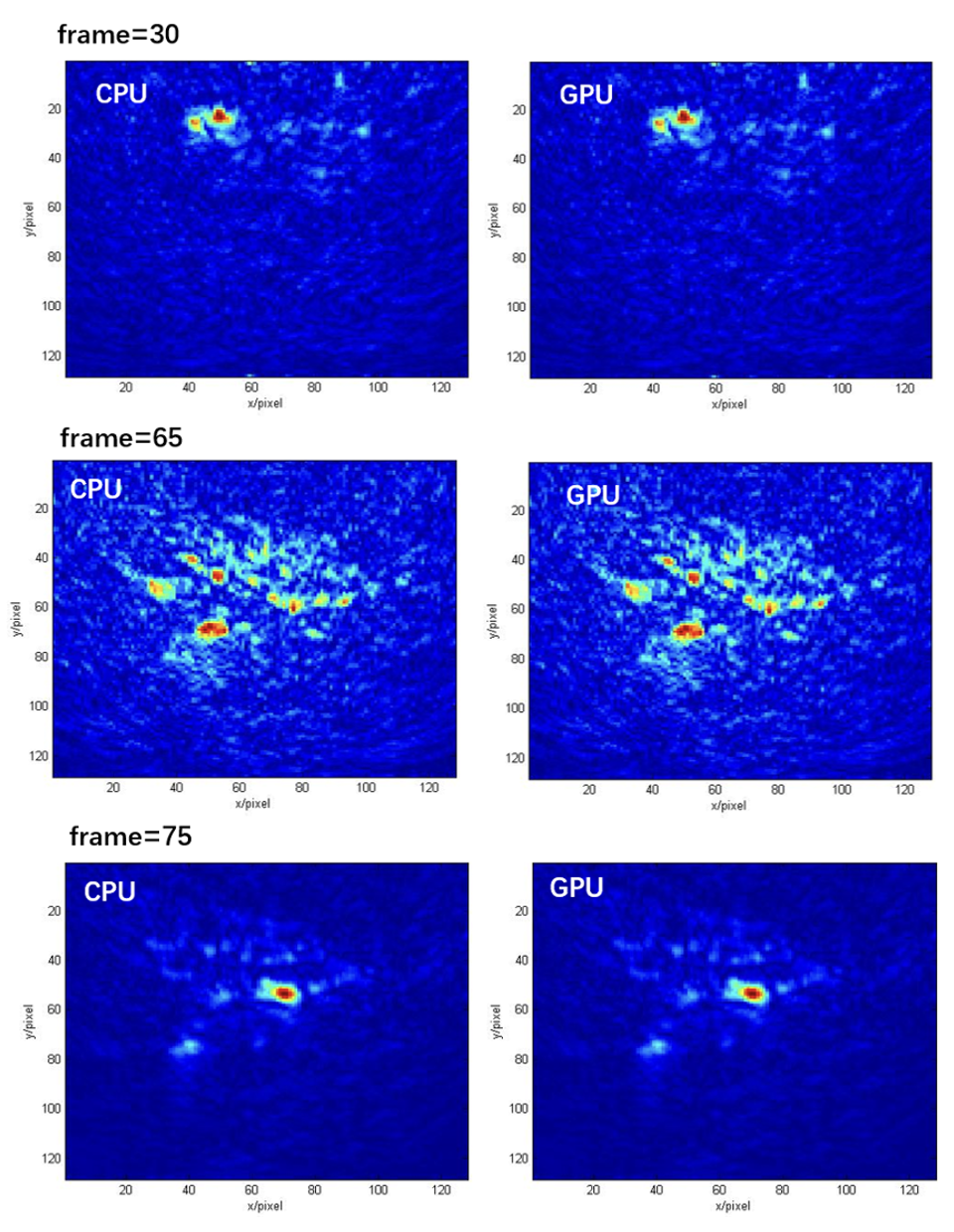}
\caption{GPU vs. CPU Imaging Performance Comparison for Frame 30, Frame 65, and Frame 70}
\label{fig:frames}
\end{figure}

The reconstruction of three frames was performed by invoking CUDA functions via MATLAB, and the corresponding run times were meticulously evaluated. The images subjected to these tests are displayed in Figures \ref{fig:frames}, with the relative processing durations comprehensively detailed in Table \ref{tab:table5}.

\begin{table}[]
\begin{tabular}{|l|l|l|l|}
\hline
\textbf{Frame} & \textbf{Image Size} & \textbf{CPU Time(s)} & \textbf{GPU Time(s)} \\ \hline
30                    & 127*127             & 118.470              & 20.062               \\ \hline
65                    & 127*127             & 118.814              & 20.039               \\ \hline
75                    & 127*127             & 117.082              & 19.892               \\ \hline
\end{tabular}
\caption{Time Consumption Comparison for the reconstruction of 3 frames between GPU and CPU}
\label{tab:table5}
\end{table}

Upon examining the three sets of reconstructed images, it is evident that the outcomes achieved using both the CPU and GPU for image reconstruction are consistent. Data from the table indicates that image reconstruction using solely the CPU requires approximately 118 seconds. In contrast, employing the GPU to expedite the reconstruction process reduces the duration to about 20 seconds, thereby achieving a speed-up factor of approximately 5.9.

\section{Conclusion}
\label{sec:conclusion}
Due to the rapid advancement of GPU hardware and the exponential increase in computing power, the application of GPUs has expanded significantly, bolstered by continual enhancements in GPU programming software. Consequently, GPU parallel computing is emerging as a potent approach for processing large datasets. This paper builds upon this concept, specifically focusing on parallelizing the labor-intensive image reconstruction process in photoacoustic imaging to expedite the imaging workflow.

Through an in-depth examination and exploration of GPU parallel frameworks, this study deepens the understanding of GPU parallel computing mechanisms. It details the conversion of traditional serial code into parallel code and the subsequent optimization of parallel photoacoustic iterative reconstruction code within the CUDA environment to boost program execution efficiency. Experimental results confirm that, in the realm of photoacoustic image reconstruction, GPUs significantly outperform CPUs in terms of image processing efficiency. Nonetheless, additional enhancements are required to fully realize the potential of this technology in clinical applications. Continuous learning and the assimilation of new knowledge are imperative, as ongoing advancements and refinements are expected to enable GPU parallel computing methods to markedly accelerate photoacoustic image reconstruction in the future, thereby promoting its extensive application in fields such as hemodynamics, clinical disease diagnostics, and pharmaceutical development.

Significant opportunities still exist for parallel optimization in the iterative processes of photoacoustic imaging. Future endeavors will aim to further refine these processes, maximizing GPU performance and optimizing data processing speeds.



\bibliographystyle{ACM-Reference-Format}
\bibliography{references}

\end{document}